\documentclass[12pt]{iopart}
\usepackage{graphicx}
\usepackage{color}

\begin{document}
\bibliographystyle{physbiol}

\title[Epigenetic Chromatin Silencing]{Epigenetic Chromatin Silencing: Bistability and Front Propagation}

\author{Mohammad Sedighi$^1$
and 
Anirvan M. Sengupta$^{1,2}$}

\address{$^1$
BioMaPS Institute, Rutgers University, 610 Taylor Road, Piscataway, NJ 08854, USA}
\address{$^2$ 
Department of Physics and Astronomy, Rutgers University, 136 Frelinghuysen Road, Piscataway, NJ 08854, USA}
 
 \ead{anirvans@physics.rutgers.edu}


\begin{abstract}
The role of post-translational modification of histones in eukaryotic gene regulation is well recognized. Epigenetic silencing of genes via heritable chromatin modifications plays a major role in cell fate specification in higher organisms. We formulate a coarse-grained model of chromatin silencing in yeast and study the conditions under which the system becomes bistable, allowing for different epigenetic states. We also study the dynamics of the boundary between the two locally stable states of chromatin: silenced and unsilenced. The model could be of use in guiding the discussion on chromatin silencing in general. In the context of silencing in budding yeast, it helps us understand the phenotype of various mutants, some of which may be non-trivial to see without the help of a mathematical model. One such example is a mutation that reduces the rate of background acetylation of particular histone side-chains that competes with the deacetylation by Sir2p. The resulting negative feedback due to a Sir protein depletion effect gives rise to interesting counter-intuitive consequences. Our mathematical analysis brings forth the different dynamical behaviors  possible within the same molecular model and guides the formulation of more refined hypotheses that could be addressed experimentally.
\end{abstract}

\maketitle

\section{Introduction}
One of the interesting aspects of developmental processes is that one could get multiple heritable cell fates without irreversible changes to the genetic information. Heritable differences in phenotype, despite having the same genetic information, goes by the name of epigenetic phenomenon.  Apart from its fundamental role in development, epigenetic effects are of great importance in certain diseases like cancer \cite{Epigenetics04, EpigeneticsAllis}.  There are many mechanisms that could lead to epigenetic effects. One of these mechanisms is transcriptional silencing.

Regions of eukaryotic chromosomes could be divided into euchromatin and heterochromatin, based on the degree of condensation during interphase, the period between nuclear divisions.
Heterochromatin refers to condensed domains where the nucleosomes, with DNA spooled around, are packed into higher order structures. Genes in this area, as opposed to genes in less condensed euchromatin areas, are not normally transcriptionally active. Therefore, the formation of heterochromatin is a way of silencing the expression of a number of adjacent genes. Furthermore, in many circumstances,  the organization of chromosomes into heterochromatin and euchromatin regions will be inherited by the new cells generated from cell divisions \cite{MBoC}. As a result, heterochromatin formation plays a crucial role in multi-cellular development by stabilizing gene expression patterns in specialized cells. One example of this is the cell type dependent silencing of Hox genes, important in development of body plans, by the Polycomb group of proteins \cite{Gilbert}.

One might distill some overarching similarities from various mechanisms proposed for silencing in diverse organisms \cite{GrewalMoazed}.
In the  general model,  there is usually a region that nucleates silencing by recruiting a silencing complex incorporating a histone modifying enzyme (figure 1). Modification of some of the lysines in the histone tails leads to binding by components of silencing complex, which, in turn, recruits further histone modifying enzymes. That is how the process propagates till it meets some boundary element (or the system reaches a stationary state due to exhaustion  of one of the components of the silencing complex).

Budding yeast, \emph{S.\ cerevisiae}, played an important role in understanding how chromatin silencing works.
In the budding yeast, there are three kinds of regions that are silenced: the telomeres, the ribosomal DNA and the silent mating type loci. There are two silent mating type loci on chromosome III, \emph{HML} and \emph{HMR}, flanking the active mating-type locus. \emph{HML} and \emph{HMR} contain copies of genes that decide $\alpha$-type and a-type identity, respectively. Information from the silent mating-type loci gets copied into the active mating locus through a recombination mediated process called mating type switching. The recombination event is initiated by a double strand break in the mating locus by the HO endonulease in haploid yeasts. The study of mating type in yeast shed light on many fundamental biological questions \cite{MBoC, MCB}.

The mechanism by which silencing nucleates and spreads  in budding yeast is relatively well investigated \cite{Rusche, Moazed} and provides a concrete example of the more general model mentioned before. It is known that the Silenced Information Regulator (SIR) proteins are the main players in gene silencing at telomeres and mating type loci in yeast. As discussed above, the model for step-wise gene silencing in S. cerevisiae (figure 1), also posits that silencing happens in two distinct steps: \emph{nucleation} and \emph{spreading}. To be concrete, let us discuss silencing at the silent mating loci. In the nucleation step, with the help of site-specific DNA binding proteins (like Rap1) and with Sir1 as a tether, Sir2, Sir3 and Sir4 will form a  Sir Complex on the nucleation site. Deacetylation of certain lysines on the neighboring histones H3 and H4, by Sir2 (a $NAD^+$ dependent histone deacetylase (HDAC)) will make binding of Sir3/Sir4 complex easier in the neighborhood of the original nucleation site. Sir3/Sir4, in turn, would recruit more Sir2. Hence, the spreading starts. More deacetylation of histone tails improves the recruitment of other  Sir proteins and formation of more stable complexes on neighboring sites.  If histone deacetylation is transferred further on, it will result  in spreading of silencing to even distal sites.  Although the nucleation step is different in telomeric silencing, the process of spreading  seems to be very similar \cite{Aparicio}.

However, in the wild type budding yeast, the regions that are silenced are, typically, always silenced. To see epigenetic effects, one needs to ``weaken" the system.
As mentioned above, Sir1 is required for efficient nucleation of silencing at \emph{HMR}/\emph{HML} loci. Experiments on {\it sir1} mutants (where the nucleation effect is defective if not absent) show that the genes at \emph{HML} loci can be either repressed or derepressed in individual cells, representing two phenotypically distinguishable cells \cite{PillusRine}. Both states are stable to small fluctuations and are typically preserved in cell divisions - there is only a small probability of transitions back and forth between states - suggesting that the system is actually in a bistable regime. 

In the systems biology community, epigenetic switches in prokaryotes have received quite a bit of attention. Multiple phenotypes are usually represented as multiple stable attractors in deterministic descriptions of the biochemical dynamics. Computational modeling of lambda phage \cite{Ptashne} has played a crucial role in the development of systems biology \cite{sheaackers, arkin}. From the response of  lac operon  in the presence of TMG    \cite{NovickWeiner, ouden, Guet} to synthetic genetic networks like the toggle switch \cite{toggle}, mathematical analysis has been an integral part of understanding such phenomena. In particular, the biological model, in each of these examples, provides a mechanism of positive feedback. However, positive feedback is not sufficient to guarantee multistability, essential for giving rise to non-trivial epigenetic states.

The crucial aspect of the analysis of the mathematical model is computing the bifurcation diagram telling us which region in the space of control parameters is actually associated with bistability. The bifurcation diagram also indicates the qualitative behavior of the system when perturbed (or mutated) in a particular manner. In contrast to  prokaryotic epigenetic switches just mentioned, modeling eukaryotic epigenetic silencing involves studying a spatially extended bistable system. Hence,  the system shows interesting  phenomena, like front propagation, allowing for a richer bifurcation diagram. 

In this paper, we introduce a mathematical model of step-wise heterochromatin silencing. A mean field description of the dynamics explains many features of the real system. Epigenetic states, in the absence of efficient nucleation, can be explained as a consequence of the existence of two stable uniform static solutions: the hyper-acetylated state and silenced states on DNA. Studying the conditions under which the positive reinforcement inherent in the proposed silencing mechanism is strong enough to give rise to bistability  is one of the main goals of our paper.   In addition, the conditions required for static fronts  will set additional constraints on the model. At the end, we propose experiments designed to test the ideas discussed here. 

\section{Methods}

\subsection{Mathematical formulation of a model of silencing}

We formulate a quantitative version of the conventional biological model of silencing \cite{GrewalMoazed, Moazed}. The  main variables involved in final equations are \emph{the local degree of acetylation} and \emph{the local probability of occupation by Sir complex}, both of which could depend on time, as well as on the position of nucleosomes on DNA, represented as a one-dimensional lattice.  We define function, $S_{i}(t)$ on this lattice, as a number between $0$ and $1$, to represent fractional number of Sir complexes at  site \emph{i}. Fractional degree of acetylation, $A_{i}(t)$, is defined in the same way too.   Writing  chemical equations,
in the mean-field treatment of the system, we get,      
\begin{eqnarray}
\frac{dS_{i}(t)}{dt}
&=&\rho_{i}(t)(1-S_{i}(t))f(1-A_{i}(t))-\eta S_{i}(t)\label{SirEq}\\
\frac{dA_{i}(t)}{dt}
&=&\alpha(1-A_{i}(t))(1-S_{i}(t))-(\lambda+\sum_{j}\gamma_{ij}S_{j}(t)) A_{i}(t)\label{AceEq}
\end{eqnarray}

In equation (\ref{SirEq}),  the first term on the right hand side is the Sir complex binding rate and the next term  is the ``fall off" rate.  The 3D concentration of ambient Sir complex at site $i$, which is generally a function of time, is denoted by $\rho_{i}(t)$. Since free Sir proteins in the environment do not form Sir complexes by themselves, this quantity actually represents a function of concentrations of all components (Sir3, Sir2 and Sir4) that are ready  to make a Sir complex on the site. For example, in the simplest case, when each protein is in low abundance,  this function would be proportional to the product of the three concentrations. However, throughout this paper we will not need to go into the details of this function. The function $f(x)$ dictates the cooperativity in Sir complex binding and should be a monotonically increasing function of $x$, $0\leq x\leq 1$. We use $f(x)=x^n$, where $n$ is the degree of cooperativity between deacetylated histone tails in recruiting Sir proteins. At last, $\eta$ is  degradation rate of  bound Sir complexes. In equation (\ref{AceEq}),  as in equation (\ref{SirEq}), on \emph{RHS}, the first term advocates  creation  and next one degradation. The parameter $\alpha$ represents the constant acetylation rate\footnote{More generally, the acetylation rate could be $\alpha(1-A_i)(1+\sigma-S_i)$ allowing acetylation of histones in silencing complex bound nucleosomes, but adding this process does not make a qualitative difference.}. In the second term, the summation accounts for the contribution of adjacent Sir complexes in deacetylation of site $i$. Since Sir complex is only capable of deacetylation of sites in its neighborhood, $\gamma_{ij}$ is assumed to be symmetric with respect to its indices and drop significantly as $|i-j|$ gets large. Finally, $\lambda$ is the rate of deacetylation from the rest of deacetylase proteins. This rate is assumed to be a constant both in time and position. 

\subsection{Generalized model including feedback from modulated transcription rate}

In the model formulated in the previous section, we allowed a certain degree of cooperativity in how deacetylated histone tails recruit the silencing complex. As we will see, this cooperativity would essential for having multistability within this model. However, the cooperativity in that particular interaction is not absolutely essential when we have other nonlinear effects in play.

One rather plausible effect is as follows. Transcription of a gene is often associated with a higher rate of acetylation of histone tails. It is believed to be one of the reasons why highly transcribed genes are hard to silence. For example, a tRNA gene, usually producing a large amount of RNA, has been found to have an important role in a silencing  boundary \cite{Kamakaka}. One might therefore imagine that silencing, which affects local transcription rates, indirectly affects the local acetylation rate. One way to model this is to introduce an additional function $g(1-S_i)$ in the local acetylation rate making it
$\alpha(1-A_{i}(t))(1-S_{i}(t))g(1-S_i(t))$. If there is no such feedback from silencing, we could have $g(y)=1$. We will consider $g(y)\sim y^{m-1}$, $m=1$ being the case of no feedback, where as the simplest models of feedback would lead to $m=2$. For a general value of $m$ (and $n$) our model now  would be given by the following equations.
\begin{eqnarray}
\frac{dS_{i}(t)}{dt}
&=&\rho_{i}(t)(1-S_{i}(t))(1-A_{i}(t))^n-\eta S_{i}(t)\label{SirEq-m}\\
\frac{dA_{i}(t)}{dt}
&=&\alpha(1-A_{i}(t))(1-S_{i}(t))^m-(\lambda+\sum_{j}\gamma_{ij}S_{j}(t)) A_{i}(t)\label{AceEq-m}
\end{eqnarray}
Thus, the nature of nonlinearity in these models is characterized by a number doublet $(m,n)$.

\subsection{Determining the nullclines and the bifurcation diagram}

One could analyze the uniform static solutions of equations and study the stability. The stationary states are obtained by solving the algebraic equations produced by setting time derivatives  to zero. We analyze first the case where \emph{available} SIR concentrations are kept at a constant level. This means $\rho_i(t)=\rho,$ a time (and position) independent number.

Dropping all $i$ indices and replacing the non-local term $\sum_{j}\gamma_{ij}S_{j}$ with $\Gamma_0S$, we can  rewrite  equations  as:
\begin{eqnarray}
\frac{dS(t)}{dt}
&=&\rho (1-S(t))f(1-A(t))-\eta S(t)\label{Uniform0}\\
\frac{dA(t)}{dt}
&=&\alpha(1-A(t))(1-S(t))g(1-S(t))-(\lambda+\Gamma_0 S(t)) A(t)
\label{Uniform}
\end{eqnarray}

Setting time derivatives to zero gives the nullclines. Taking derivatives of nullclines with respect to $A$ (or $S$) and setting these derivatives 
equal to each other provides us with the condition for the saddle-node bifurcation determining the boundary of bistability region. Finally we write $\alpha$ and $\rho$, at this boundary, parametrically in terms of $S$. For example, for the $(1,n)$ models, where $n$ is the degree of cooperativity ($f(x)=x^n$), we have:
\begin{eqnarray}
\alpha
&=&\frac{(\lambda+\Gamma_0 S)^2}{(1-S)[(\Gamma_0+\lambda)(n-1)S-\lambda(1-S)]}\\
\rho
&=&\frac{\eta S}{(1-S)}\left(\frac{n(\Gamma_0+\lambda)S}{(\Gamma_0+\lambda)(n-1)S-\lambda(1-S)}\right)^n
\label{Nullclines}
\end{eqnarray}

 \subsection{Study of non-uniform solutions}
 
In the following subsections, we go beyond analyzing the stable uniform solutions. In the region of parameter space where the system is bistable, it is possible to study how fronts between a silenced region and an unsilenced region move. In a system with a well defined free energy function, the average motion of a front or interface is determined by the difference of free energies of the two states across the front. Deterministically speaking the lower free energy state (usually called the stable state) would invade the metastable state with higher free energy. At the points where the two free energies are the same, the average front velocity is zero. Although, in  non-equilibrium systems, like the one at hand, there is no useful free energy to be defined, one might still explore the region of parameter space where silenced state invades the unsilenced ones and vice versa (and the line in between where the front becomes stationary).

We study the motion of boundary between the two stable phases in the bistable  parameter region both in the current discrete model and in a local continuum version of the model where the lattice is replaced by a continuous 1d system. 

\subsection{Formulation of the continuum version of the model}

In the continuum formulation, we replace the index $i$ by a continuous variable $x$, $x=ia$, where $a$ is the lattice spacing. The variables $A_i,S_i$ become functions of $x$, namely $A(x), S(x)$. If we define $\gamma(z)$ to be $\gamma_{ij}/a$, where $z=(j-i)a$.
When $a$ tends to zero the expression $\sum_j\gamma_{ij}S_j$ becomes $\int dz \gamma(z)S(x+z)$. By Taylor expanding $S(x+z)$ in $z$ and keeping up to the second order term (note that, this approximation is valid since $\gamma(z)$ drops to zero as $z/a$ increases), we get
the local continuum versions of  equations (\ref{SirEq}) and (\ref{AceEq}) which become
\begin{eqnarray}
\frac{\partial S(x,t)}{\partial t}&=&\rho(1-S(x,t))f(1-A(x,t))-\eta S(x,t)\\
\frac{\partial A(x,t)}{\partial t}&=&\alpha(1-A(x,t))(1-S(x,t))g(1-S(x,t))-\nonumber\\
& &\left(\lambda+\Gamma_{0}S(x,t)+\Gamma_{2}\frac{\partial^2 S(x,t)}{\partial x^2}\right)A(x,t)
\end{eqnarray}
where $\Gamma_{0}=\int\gamma(z)dz$ and $\Gamma_{2}=1/2\int\gamma(z)z^2dz$. 
For each set of parameters,  there is a front velocity \cite{AronsonWeinberger, CrossHohenberg, CompCellBio}, $c$, for which there is only ``one" (or none) continuous solution that represents a transition between the stationary states representing euchromatin and silenced heterochromatin.

\subsection{Computing the zero velocity line in the continuum model}
The analysis of the continuum system follows the standard route \cite{AronsonWeinberger, CrossHohenberg, CompCellBio}.  Assuming $A(x,t)=A(x-ct)$ and $S(x,t)=S(x-ct)$,  and using $u=x-ct$: 
\begin{eqnarray}
c\frac{\ dS(u)}{\ du}+\rho(1-S(u))f(1-A(u))-\eta S(u)&=&0\label{SirEqContSimple}\\
c\frac{\ dA(u)}{\ du}+\alpha(1-A(u))(1-S(u))g(1-S(u))-& &\nonumber\\
\left(\lambda+\Gamma_{0}S(u)+\Gamma_{2}\frac{\ d^2 S(u)}{\ du^2}\right)A(u)&=&0\label{AceEqContSimple}
\end{eqnarray}

The analysis of problem for $c=0$ is considerably simpler, since equation (\ref{SirEqContSimple}) turns out to be an algebraic equation allowing $A$ to be expressed in terms of $S$, namely, $$A(S)=1-f^{-1}(\frac{\eta S}{\rho (1-S)}).$$ We define a potential $V(S)=-\lambda S-\Gamma_0 S^2/2+\alpha \int dS (1-A(S))(1-S)g(1-S)/A(S)$, so that other equation, namely equation (\ref{AceEqContSimple}), could be written as
\begin{equation}
\Gamma_{2}\frac{\ d^2 S}{\ dz^2}=\frac{\ dV(S)}{\ dS}
\end{equation}
The values of parameters, for which the potential $V(S)$ has two local minima  with equal potential values, correspond to existence of a zero velocity front. Note that we could use this potential only to describe zero velocity fronts, and not for the general traveling solution.

\subsection{Computing the zero velocity line in the discrete model}

Rewriting nullclines for discrete model:
\begin{eqnarray}
S_{i}
&=&\frac{\rho(1-A_{i})^n}{\eta+\rho(1-A_{i})^n}\label{Uniform1}\\
A_{i}
&=&\frac{\alpha(1-S_{i})^{m}}{\lambda+\sum_{j}\gamma_{ij}S_{j}+\alpha(1-S_{i})^{m}}
\label{Uniform2}
\end{eqnarray}

we checked whether above equations allow a zero velocity barrier as a fixed point. For each pair of $\alpha$ and $\rho$ inside the bistable region, starting with an array of $i_{max}=100$ sites, 
with fixed boundary conditions and initial condition of half recursive sites silenced/deacetylated and the remaining unsilenced/acetylated (with values of $S$ 
and $A$ obtained from fixed points of corresponding \emph{uniform} nullclines, obtained by setting the right hand sides of the equations (\ref{Uniform0}) and (\ref{Uniform}) to zero), we applied equations (\ref{Uniform1}) and (\ref{Uniform2}) respectively. By iteration we let 
the system evolve until it reached its fixed point (within an error of 0.01\%). 

\subsection{Dynamics with Sir protein depletion} 

Finally, we consider the effect of $\rho(t)$ not being constant. We model the limited supply of Sir proteins by putting a constraint on total number of Sir complexes, in solution and on the DNA.
Going back to original equations ( \ref{SirEq}) and (\ref{AceEq}), we need to replace the constant $\rho_{i}(t)$ by a $\rho(t)$ which is given as follows:
\begin{equation}
\rho(t)=(S_{total}-\sum_{k}S_{k}(t))/V
\end{equation}
where $S_{total}$ is total number of functional Sir complexes in the system, which is constant and $V$ represents the average volume. 

\section{Results}
\subsection{Bifurcation analysis of the model of silencing}

One could analyze the uniform static solutions of equations and study the stability. The stationary states are obtained by solving the algebraic equations produced by setting time derivatives  to zero. We analyze first the case where \emph{available} SIR concentrations are kept at a constant level. This means $\rho_i(t)=\rho,$ a time (and position) independent number. One can see that, for $f(x)=x^n, g(y)=1$, depending on chemical parameters, one can get either one or three fixed points (figure 2), provided $n>1$. The three fixed point case  always includes two stable points enclosing the other unstable saddle point, so the system is actually in a bistable state as could be seen by local analysis (figure 3). One of the two states has low acetylation and higher chance of repression, while the other state has a high degree of acetylation and higher chance of derepression (figure 3). 

The bifurcation diagram, indicating regions in the parameter space of $\rho$ and $\alpha$ leading to monostability and to bistability, is shown in figure 4 (solid lines). Note that the critical point of this bifurcation is at low availability silencing factors coupled with low rate of acetylation. This feature will have an important implication when we consider mutants lacking particular acetyltransferases. 

The constraint that $n$ needs to be greater than one suggests that in the simplest models (namely the $(1,n)$ models), we need a certain degree of cooperativity in recruitment of Sir complexes. Currently there in not much evidence for or against such a cooperative effect from deacetylated histone side chains. In the next subsection, we point out that in a more generalized model, such an effect is not essential.

\subsection{Cooperativity versus feedback from modulated transcription rate}

We discussed $(1,n)$ models in the previous section and found that we need $n$ to be greater than one for these subclass of models. Therefore, we analyzed the more general model with transcriptional feedback, described by equations (\ref{SirEq-m}) and (\ref{AceEq-m}).  We also found that if we let $m=2$, we could get bistability with $n=1$ (results not shown). Moreover, the structure of the bifurcation diagram is essentially the same. Since both kinds of models, those with Sir binding depending strongly nonlinearly on the degree of deacetylation, as well as those where the effect of silencing on local transcription feeds back on the acetylation rate, provide qualitatively similar results for many of the predictions to be made in later sections, we will continue to show the results of the $(1,n)$ models, fully keeping in mind that there is a broader class of models leading to the same qualitative predictions.

 \subsection{Nonuniform solutions and front propagation}

For the purpose of this paper, we would focus on the part of the parameter space where the front velocity is zero.   The zero velocity line is represented by the  `dashed' line in figure 3. This line divides the bistable region into two parts. In the upper half, a front would move abolishing silencing, whereas, in the lower half, the movement would spread silencing. 

We have also studied the discrete model directly. As expected, discrete model gives a band in the parameter space for front propagation failure \cite{Keener}. The boundaries of the band are represented by the  dotted lines in figure 3.
This band shrinks to the zero velocity line as one takes the continuum limit. One might ask which of these descriptions are closer to reality. If we count each nucleosome as a unit and expect one silencing complex per nucleosome, then that provides us with a natural lattice spacing. However, the nucleosomes are not quite static. They could move around or disappear (if the histone octamer falls off DNA). If the time scale of nucleosome dynamics is much slower than that of the silencing process, then we are justified in taking the nucleosome array as a lattice to operate upon. If the time scales are the other way around, we might average out the nucleosome fluctuations and get an effective continuum description. The truth probably is somewhere in between, leading to a fuzzy region of low front mobility crossing over to high front mobility regions above and below in the bifurcation diagram.

\subsection{The role of finite supply of Sir proteins}                    

The previous discussion assumed that the available ambient concentrations of Sir proteins were constant, reflected in $\rho$ being held constant. We could use our insights, into the bifurcation diagram, to infer what would happen if the total number of Sir proteins (the sum of those in solution and those bound to DNA) were fixed. This is particularly interesting in the bistable region. 

Our treatment is very similar to studying phase equilibrium with a fixed number of particles. For example, consider a liquid gas mixture at a constant temperature in a fixed volume with fixed number of particles, and imagine that there is an interface between the two states.  The interface moves, and the fractions of particles in the different states  change till  chemical potential of the two states become equal. Under  this final condition, the interface does not move anymore, apart from thermal fluctuations around the average position. As we noted before, in our problem, we may not define a free energy, but we could indeed talk about average movement of interface between two states, namely the front, and the condition under which the interface stops moving.

Depending upon the size of the silenced region, one would get interesting titration effects in this model.  Suppose we are in the bistable region of the parameter space and start from the locally stable uniform unsilenced solution. Let us ask, what happens if we nucleate a small region of silencing, say, by tethering a protein that recruits the silencing factors locally. If we are in the upper half of the bistable region (Region I in figure 4), high acetylation rate makes difficult for silenced region to spread into the unsilenced region. In that case, in the deterministic model, silencing is going to remain localized around the region of recruitment.

 Now, if the acetylation rate, $\alpha$, is tuned down, say, by knocking off an acetyltransferase;  we go into the region  where the silencing can spread into unsilenced DNA (Region II in figure 4). So, naively, we expect much more silencing. 
However,  since $\rho$ is no longer fixed, as silencing spreads, $\rho(t)$ reduces. At this point one of the two following things could happen. The silencing could stop at special boundary elements on DNA where some other process stops the spread of silencing \cite{BiBroach, Kamakaka}. Alternatively, the front could stop because $\rho(t)$ reduces enough to reach  a point on  bifurcation diagram where the propagation velocity is zero. Thus, in this case, the effect of reducing $\alpha$ is to effectively reduce $\rho$ as well, so that the system always stays on the zero velocity region. Note that, if there are more than one region in DNA where silencing spreads by the same mechanism and if at least one of these regions does not possess a boundary element, then  we are led to the same situation, namely  $\rho$ reducing enough to stop front movement. We will explore the biological consequence of this observation next.

\subsection{Predictions from the Model}

The bifurcation diagram presents  a classification of  qualitatively different kinds of dynamics  possible within the model. It provides us with a more precise vocabulary for discussing qualitative consequences of alternative models. Combining this with experimental facts, we should be able to place the wild type yeast and various mutants in this diagram. Outside the bistable region, the dynamics decides a self-consistent level of silencing. For instance, for parameters chosen from above the bistable region in figure 4, recruitment of silencing complex at one place only affects a small region around it, with the effects dying off exponentially with distance from the nucleation center. The upper part of the bistable region, with higher values of $\alpha$ (Region I), is not qualitatively very different in that regard.
The only difference comes in, when one considers stochastic dynamics, which allows for occasional formation of silencing in the whole region. 

In the lower half of the region, Region II, nucleation leads to spreading. This is the region where the naive expectation from the popular biological model matches the results of mathematical analysis. We have argued, that under some conditions, the dynamics of Sir depletion would lead the systems starting in  this region into the border of the two regions (zero front velocity curve, figure 4). A locus of DNA, described by parameters of Region II, could possibly see non-specific silencing induced by stochastic nucleation of silencing.

In this bifurcation diagram, where is the point corresponding to silencing dynamics in silent mating loci in wild type yeast? The fact that the silent loci in the {\it sir1} mutants could be in either state, suggests that one is in the region allowing bistability. A tougher question to answer is which part of the bistable region it is in. In Region II, there is quite some chance of getting undesirable non-specific silencing.  As one moves away from the cusp point (or critical point, figure 4), the rate of stochastic switching to silenced state can in principle become very small for regions close to the zero velocity line. However, it is really a matter of numbers. Given that the genome size is large one needs the probability of non-specific nucleation to be small enough to prevent occurrences of inadvertent local silencing. On the other hand, in Region I, typically, the silencing spreads very little from the nucleation center, unless the system is close to the cusp point. In fact, if one defines a length scale by how far the effect of silencing of local nucleation spreads, that length scale diverges exactly at the cusp point. So the system could also operate at a point where this length scale is large but not too close to the critical point where the switching rate is high.

The dynamics represented in the popular cartoon model of silencing, reviewed in \cite{GrewalMoazed}, 
corresponds to the behavior in Region II. Such models come with explicit  requirement of boundary elements to stop the spreading. On top of that, as mentioned before, there should be an argument why chance nucleation in somewhere else in the genome does not cause spontaneous non-specific silencing. However, if the system is in Region I, then one could observe a reduction in silencing with increasing the distance from the nucleation  center, namely the silencer. Such a claim has been made by some researchers  \cite{BroachPNAS}. 

Perhaps the most crucial result of our analysis is the elaboration of these two distinct possibilities within the same molecular model. It may not be crucial to decide in favor of one or the other scenario, given the likelihood of spatial inhomogeneity of the parameters. If the value of effective parameters like $\alpha$ varies in space, different sections of chromosome can demonstrate different silencing behaviors depending on what regions of bifurcation diagram they corresond to. Therefore more investigations are required to decide where on bifurcation diagram the whole or different sections of wild type chromosome are actually located. As a result, the qualitative picture that our model suggests sheds more light on the direction of future investigations in this matter.

Independent of where the wild type yeast is located in parameter space, we could discuss the consequences of lowering the acetylation rate as it happens in, say, the {\it sas2} mutant  \cite{Horikoshi, Grunstein}. We argued that if there are fronts of silencing that are not pinned down by boundary elements somewhere in the genome, then our argument about $\rho$ (Sir binding rate) reducing and  moving the system back to the zero velocity line/region, applies. This is indeed a possibility in yeast. Although the silent mating loci have well defined boundary elements, the same may not be true of all the telomeric regions. This result might explain certain counterintuitive features of mutants of certain genes like {\it sas2 } which code for acetyltransferases.  
 If the reduced acetylation rate in {\it sas2} mutant is within a certain range, the system will {\it always} flow back closer to the cusp point at the tip of the bistable region thanks to Sir titration effects. Near the cusp point, the degree of silencing changes very sharply with the changes of Sir availability (see figure 5). We believe that the resulting system becomes extremely susceptible to cellular noise and would display a wide distribution of expression. Thus, as opposed to  the naive expectation that {\it SAS2} deletion will just make every thing more transcriptionally silent, one should find individual cells that show good expression from the ``silent" loci.
We speculate, whether this is the reason why the {\it SAS genes} may have been picked up in an assay looking for defects in silencing. Recent single cell measurements observation for GFP expression from {\it sir1sas2} cells show a wide but unimodal distribution of expression in a cell population, where as {\it sir1} cell population show bimodal distribution, characteristic of epigenetic  states \cite{SingleCell}.

Another simple consequence of the bifurcation diagram is that one could say qualitative things about the epigenetic switching rate in different parts of the the bifurcation diagram. For example, we expect the switching rate to get faster near the cusp point. We expect as the level of Sas2 is lowered continuously,
we will see  a rise in switching rate, as  the system would move toward the cusp point.

\section{Discussion}

We have formulated a mathematical version of the model of silencing and computed the bifurcation diagram of the system. This diagram is consistent with several observations about mutants. It is, in principle, possible to explore the whole two dimensional control parameter space experimentally. For example, one could study single cell  fluorescent protein expressions from reporters in \emph{HML} and in \emph{HMR} while modifying $\rho$ by regulating Sir proteins, and modulating $\alpha$ via changing the level of Sas2. 

In addition to the  {\it sas2} mutant, which we discussed extensively, one  of the mutants that we want to understand is {\it dot1}. Part of the reason to study this mathematical model is the apparent paradox: if the Sir2,3,4 system itself can propagate further from region with stochastic nucleation of silencing, why  many other regions, not contiguous to silencing at nucleation sites, do not show occasional heritable silencing? In fact, a screen for high copy disruptors of telomeric silence \cite{Gottschling}, produced, among others, a gene called {\it DOT1} whose deletion cause nonspecific silencing.  Understanding how Dot1 affects silencing requires us to consider additional states like methylation of histones \cite{Dot1}. Based on our preliminary study of a full model of the system with additional states it seems that the simpler model studied in this paper, with some change of parameters, could effectively capture the effect of Dot1. This is one future direction that we are pursuing.

We have touched upon the effects of noise but have not explicitly made a stochastic model. Fluctuation in bio-molecular networks has been the subject of  many
research  activities recently \cite{rao}.  To analyze single cell data, one needs to know not only how the deterministic model behaves but also how noise in various quantities affects expression. A stochastic version of the model, a lattice model with local states of acetylation, and Sir occupancy, could be studied by direct simulation.  However, as seen in studies of yeast gene expression \cite{euknoise, OShea}, extrinsic noise, equivalent to fluctuations in the parameters themselves, often dominates over intrinsic fluctuations of the processes described here with fixed parameters. Hence, to study this properly, we will need to add a free parameter each characterizing the slow noise in the control parameters $\rho$ and $\alpha$ for modeling the effect of cell to cell variation of Sir proteins and acetyltransferases. However, we need to  be careful to avoid overfitting the data. Another interesting direction involves modeling of noise induced transition between epigenetic states. 

We finally mention two issues not dealt at all within this paper that needs further attention. One is that our model of DNA, as a one dimensional system, may be called into question if the heterochromatin formation happens very fast (compared to the speed with which silencing spreads), making the DNA fold up into higher order organization quickly. The other interesting issue is inheritance of silencing. Could we have our model capture inheritance in a coarse grained manner, or do we stand to gain something by modeling the probable silencing of duplicated DNA explicitly? Of course, for any biological model, there are many ways of making it more realistic. However, not many of these `improvements' change the qualitative properties of the bifurcation diagram. We believe our model includes enough features of the biological phenomena to be a good starting point for more refined discussion of the qualitative behavior of this system.

\section{Conclusions and Outlook}

After submitting this manuscript for publication, we noted a recently published  paper \cite{Dodd} on a model for chromatin silencing in {\it S. pombe}. An interesting point noted in that paper is that if there are more than two states of the histone modification, it is possible to arrange the parameters system to have stable epigenetic states without  cooperativity in any other process. Given that, in {\it S. cerevisiae}, particular histone methylations play a role in activation \cite{Dot1, Madhani}, it is worth looking at their role in bistability.

In biological context, the discussion of nucleation of silencing is mostly focussed on the process of assisted nucleation. The propensity of the system (or the lack thereof) to have spontaneous nucleation has not received as much attention. For example, finding mutations which enhance the probability of spontaneous nucleation would be of great interest. The study on disruptors of telomeric silencing has possibly already unearthed some of these mutants \cite{Gottschling} in {\it S. cerevisiae}. However, there could be much more to the mechanism of control of silencing. This is one place where theoretical studies could possibly suggest what specific signature to look for,  spurring on further experimental study. 

The issue of spontaneous nucleation is intimately tied to the question of noise induced switching of epigenetic states. In the context of epigenetic switches involving feedback through regulatory proteins, there has been much theoretical work done \cite{AurellSneppen, AurellBrown, Roma, Hornos, KimWang}. However, for chromosomal epigenetic mechanisms one faces a new class of problems. A crucial issue is whether the action of the histone modifying enzyme is very local and the silencing spreads nucleosome to nucleosome along the length of DNA or is so non-local enough so that every nucleosome in the locus affects each other and that we are essentially in a mean-field regime. Answers to these questions are currently unknown.

Our understanding of the role of epigenetic effects in  the control of  human embryonic cell fate is going through a revolution at this moment \cite{YoungCell, LanderCell}. As the key molecular players and their interactions with the chromatin become well specified, we need a sophisticated model to understand  cellular memory and its location specificity in the genome. Our experience with modeling the Sir-dependent silencing in yeast, and our ability to make refined  predictions would be an invaluable guide in dealing with the complexities of metazoan development.

\ack

We acknowledge many informative discussions with James Broach, Bradley Cairns, Marc Gartenberg, Vincenzo Pirrotta and John Widom. One of the authors (A.\ S.) thanks Daniel Gottschling for explaining the role of {\it DOT} genes.  We are grateful to Vijayalakshmi Nagaraj and Andrei Ruckenstein for their comments on the manuscript. Anirvan Sengupta was supported by NGHRI grant R01HG03470. Mohammad Sedighi was partially suppoerted by the the NIH workforce training grant R90DK071502. This research benefitted from several visits to KITP programs which were supported in part by the National Science Foundation under Grant No. PHY99-07949.

\section*{Glossary}

\renewcommand{\descriptionlabel}[1]%
         {\hspace{\labelsep}\textit{#1}}
\begin{description}

\item[Epigenetics.] The study of heritable changes in a gene's functioning that occur without irreversible changes in the DNA sequence.

\item[Chromatin.]  The complex of DNA and protein making up chromosomes

\item[Euchromatin.] A lightly packed form of chromatin, often actively transcribed.

\item[Heterochromatin.]  A tightly packed form of chromatin, usually with limited transcription.

\item[Gene silencing.]  Switching off a gene by an epigenetic mechanism.

\item[Sir proteins.] A set of budding yeast proteins involved in {\bf S}ilent mating type {\bf I}nformation {\bf R}egulation.
\end{description}

\pagebreak     
\section*{Figure Captions}
 \renewcommand{\labelenumi}{\textbf{Figure \arabic{enumi}.}}
  \renewcommand{\labelenumii}{\textbf{\Alph{enumii}.}}
\begin{enumerate}

\item  A model for nucleation and spreading of silencing in budding yeast, \emph{S. cerevisiae}.

\item  The intersections of two null-cline curves, one representing  static ``SIR binding" (dashed line) and the other one  static ``Acetylation"  (solid line), show fixed points of the system in the uniform regime. All graphs are plotted from static solutions of equations \ref{Uniform0} and \ref{Uniform} , with $g(y)=1$, $\Gamma_0/\alpha=0.6$ and $\lambda/\alpha=0.15$. 
   \begin{enumerate}
\item When there is no deacetylation cooperativity in Sir complex binding, $f(x)$ is linear and there is no bistability (only one fixed point solution). $\eta/\rho=0.05$ for this graph. 
\item Bistability is a product of cooperativity. All parameters are the same as in graph A, except that $f(x)=x^4$
\item Ambient Sir complex concentration acts as a switch for the bistable system. This graph shows how low concentration of Sir Complex pushes the system towards euchromatin solution. $f(x)$ and all parameters are the same as in graph B, except that $\eta/\rho=0.1$
\item High concentration of Sir Complex. This time, the system is pushed towards heterochromatin. $f(x)$ and all parameters are the same as in graph B, except that $\eta/\rho=0.008$.
  \end{enumerate}
  
\item The bifurcation diagram in acetylation rate, $\alpha$ and  silencing factor binding rate, $\rho$.

\item Changes in $\rho$ in response to decreasing of $\alpha$, when the total supply of Sir complexes are limited.

\item Sir occupancy as a function of Sir availabilty. The S shaped curve indicates multiple solutions as is common in a cusp bifurcation. 

\end{enumerate}
 
\pagebreak      

\begin{tabular}{l}
\includegraphics[width=150mm]{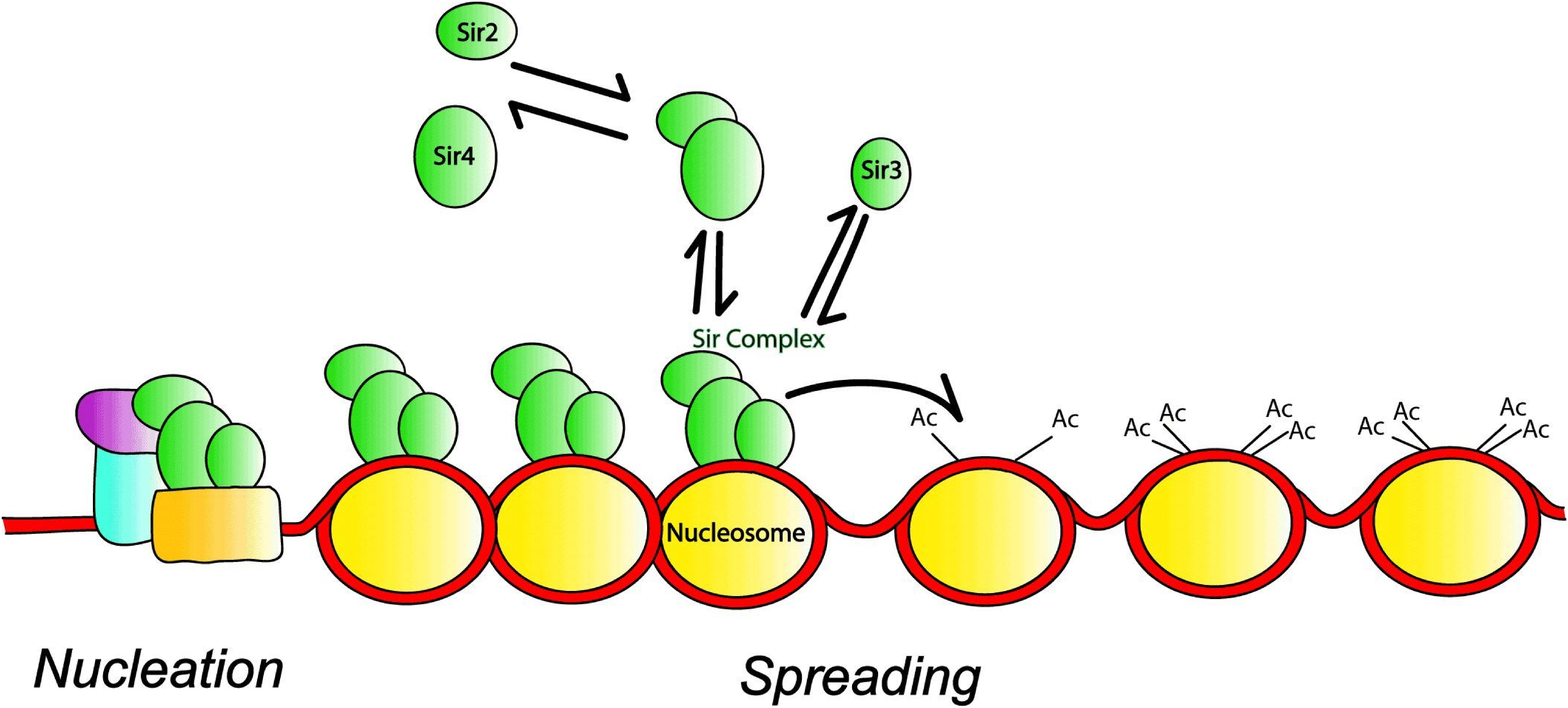}\\
\parbox{100mm}{\textbf{Figure 1. }}
\end{tabular}

\begin{tabular}{c}
\includegraphics[width=160mm]{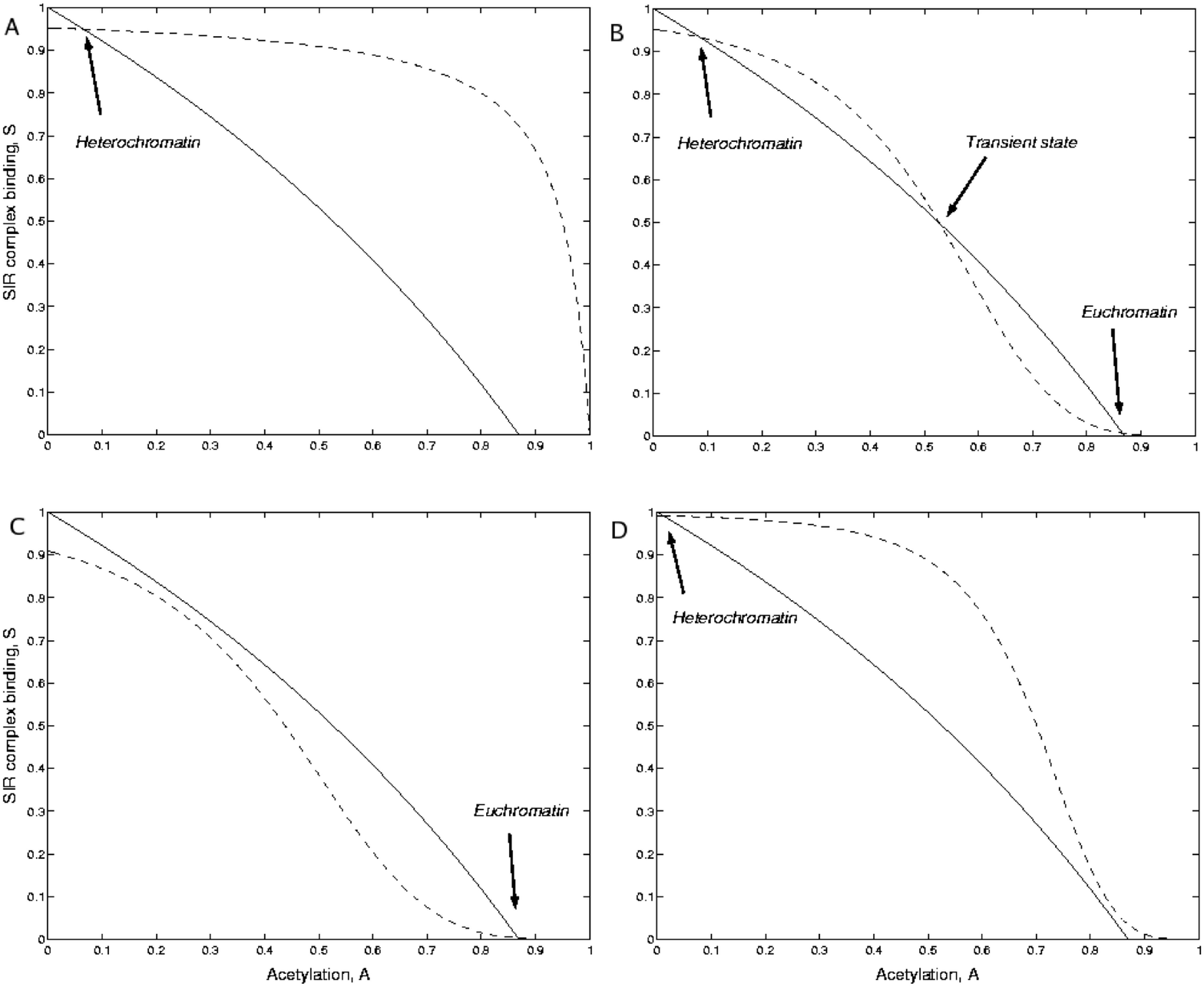}\\
\parbox{140mm}{\textbf{Figure 2. }}     
\end{tabular}

\begin{tabular}{c}
\includegraphics[width=140mm]{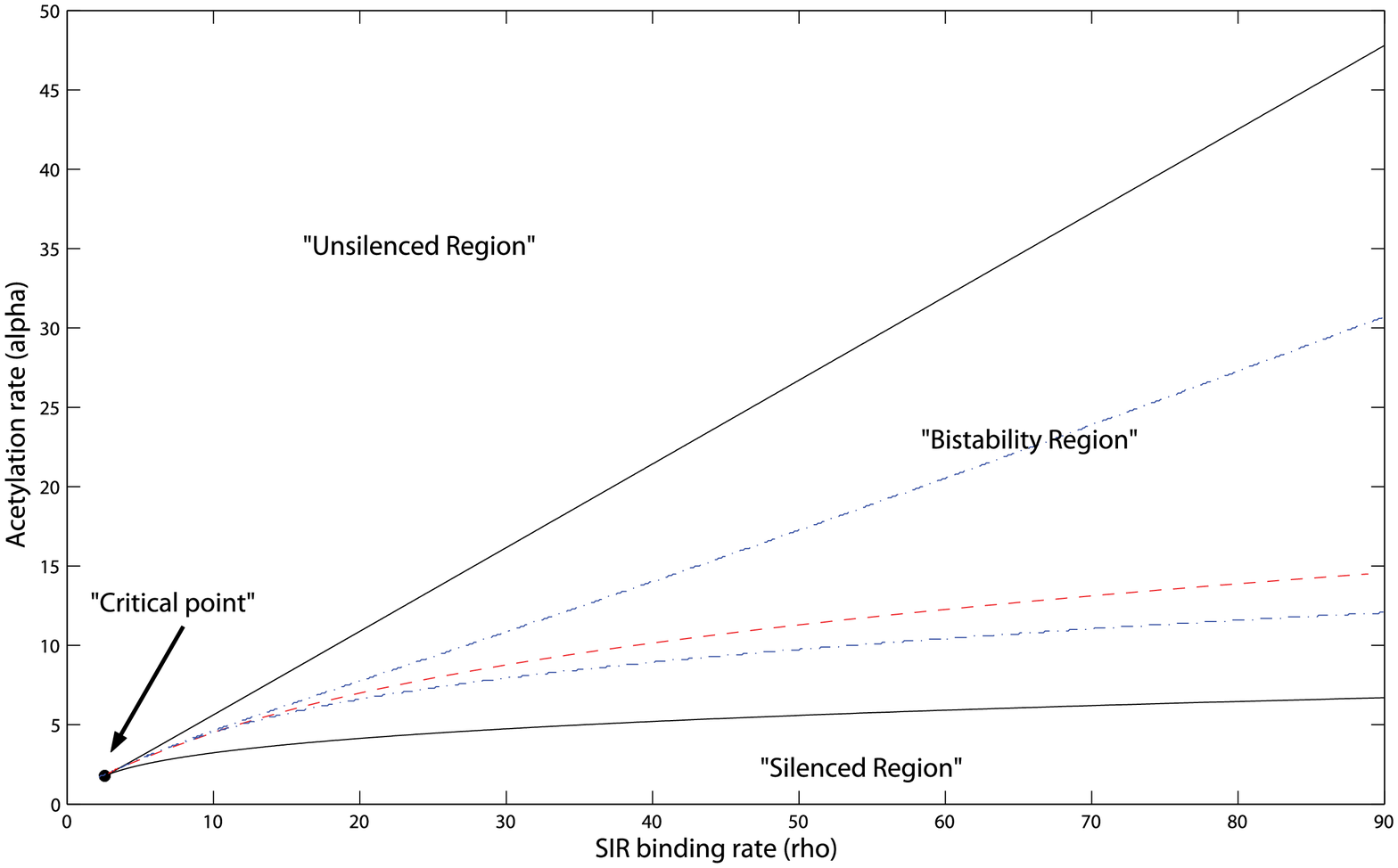}\\
\parbox{100mm}{\textbf{Figure 3. }}
\end{tabular}

\begin{tabular}{c}
\includegraphics[width=140mm]{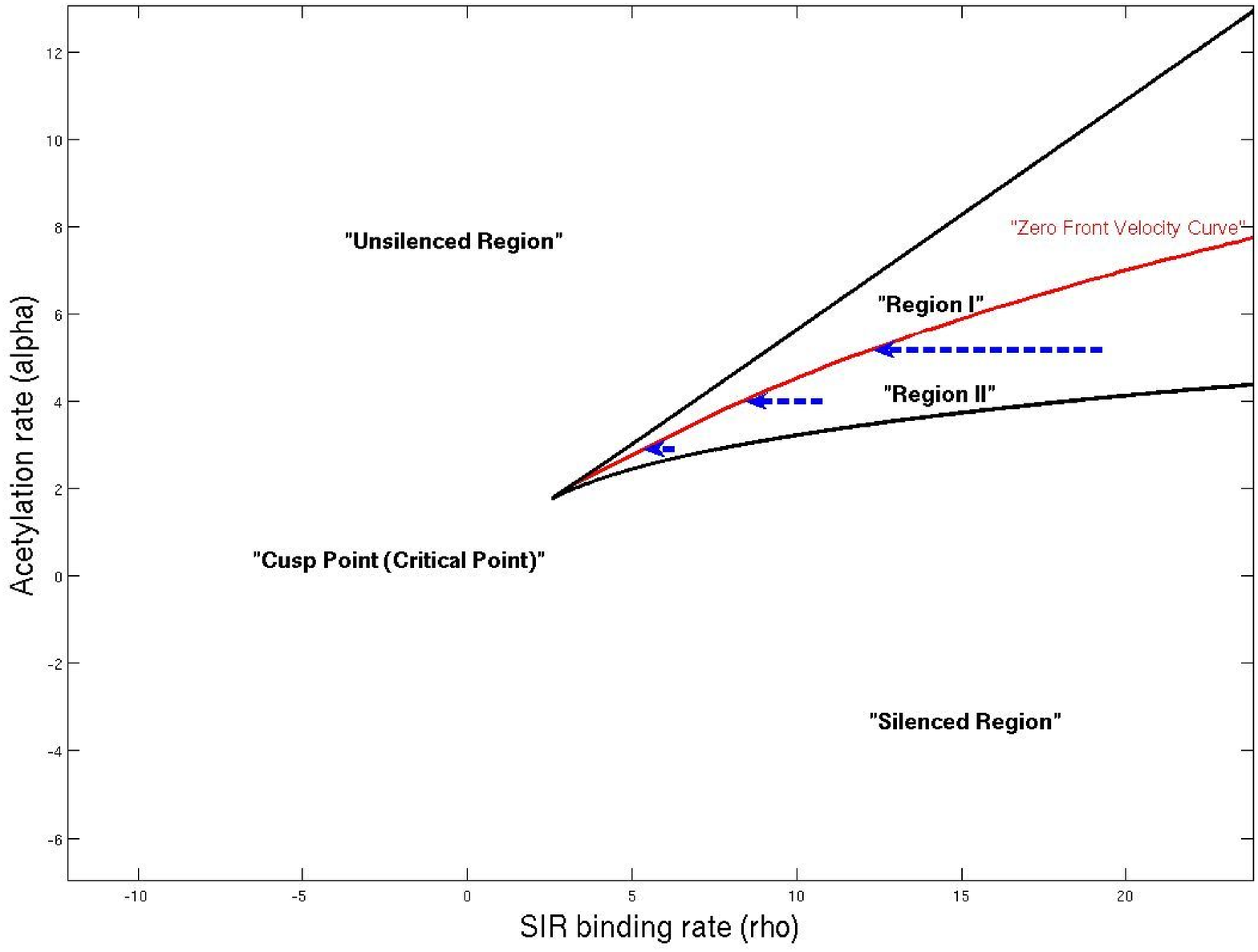}\\
\parbox{100mm}{\textbf{Figure 4. }}
\end{tabular}

\begin{tabular}{c}
\includegraphics[width=140mm]{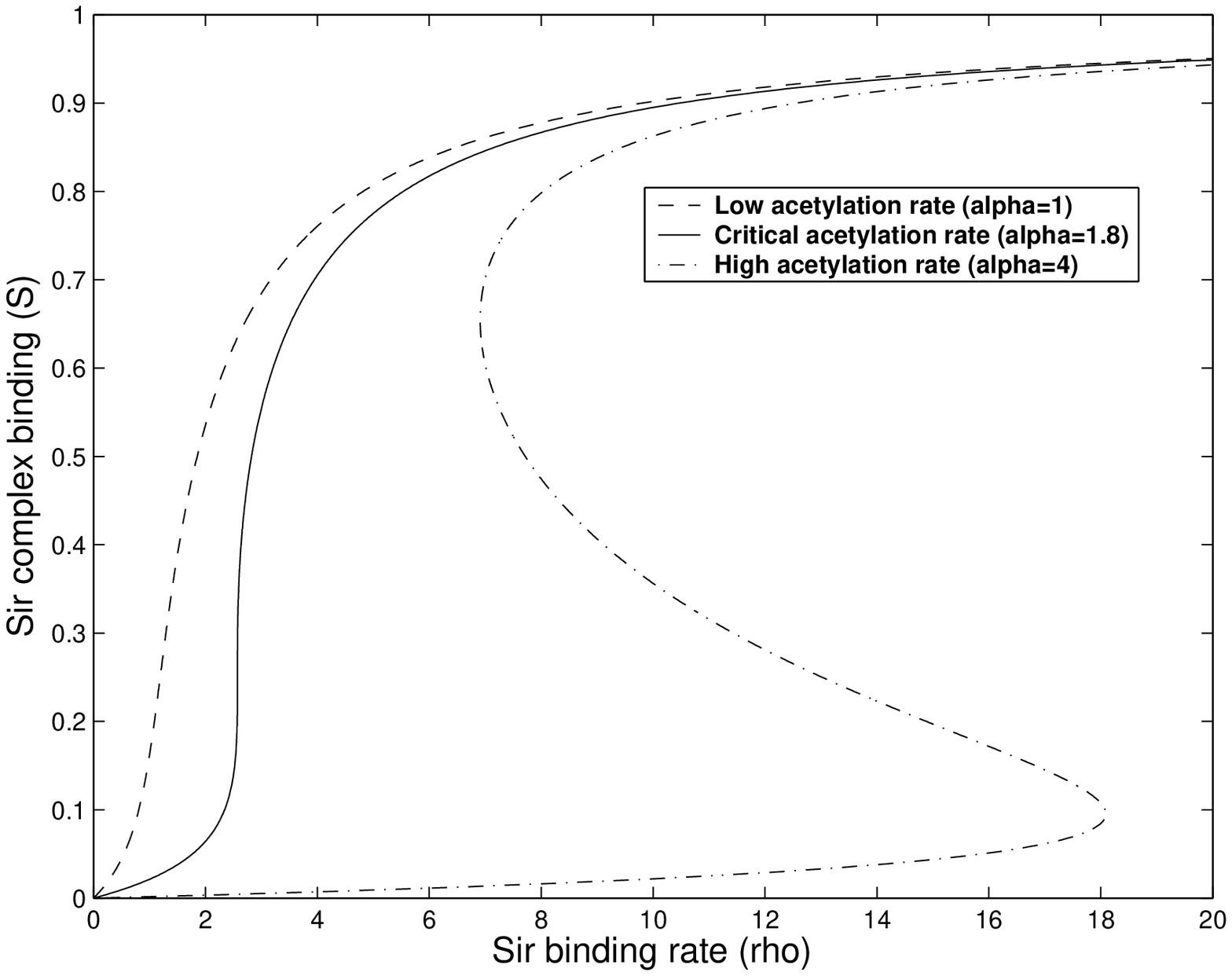}\\
\parbox{100mm}{\textbf{Figure 5. }}
\end{tabular}
\end{document}